# Representing Rod-Shaped Protein 3D Structures in Cylindrical Coordinates*

**Srujana Cheguri[1] and Vicente M. Reyes, Ph.D.[2], [3]**

(*, M. S.  Thesis, part 2; [1], M. S. student; [2], thesis advisor)
[3], E-mail:  **vmrsbi.RIT.biology@gmail.com**

Submitted in partial fulfillment of the requirements for the **Master of Science** degree
in Bioinformatics at the Rochester Institute of Technology

**Srujana Cheguri**

Dept. of Biological Sciences, School of Life Sciences
Rochester Institute of Technology
One Lomb Memorial Drive, Rochester, NY 14623

November 2011

11-1-2011

# Part 1: Size-independent quantification of ligand binding site depth in receptor proteins. Part 2: Representing rod-shaped protein 3d structures in cylindrical coordinates

Srujana Cheguri

# *Part 1: Size-Independent Quantification of Ligand Binding Site Depth in Receptor Proteins*

# *Part 2: Representing Rod-Shaped Protein 3D Structures in Cylindrical Coordinates*

Srujana Cheguri

**Approved:** ________________________________

**Vicente Reyes, Ph.D.**
*Thesis Advisor*

________________________________

**Gary Skuse, Ph.D.**
*Committee Member*

________________________________

**Paul Craig, Ph.D.**
*Committee Member*

*This thesis is dedicated to my beloved family, to my parents for their never-ending encouragement, love and confidence in me and to my brother for his motivation and guidance.*

# ACKNOWLEDGEMENTS

I feel it as a unique privilege, combined with immense happiness, to acknowledge the contributions and support of all the wonderful people who have been responsible for the completion of my master's degree. The two and half years of graduate study at RIT has taught me that creative instinct, excellent fellowship and perceptiveness are the very essence of science. They not only impart knowledge but also place emphasis on the overall development of an individual. I am extremely appreciative of RIT, especially the Department of Biological Sciences (Bioinformatics Option) in this regard. I owe it to my mentors at RIT to what I am today.

I would like to express my deepest gratitude to my thesis advisor, Dr. Vicente Reyes who continually encouraged and guided me during the course of my thesis work. I would also like to express my appreciation to my committee members Dr. Gary Skuse and Dr. Paul Craig for their valuable guidance, timely help and support. A special thanks to Nicoletta Bruno Collins for all the academic formalities that was needed to be done.

I would like to thank all professors, mentors, family and friends who have helped me scale heights and achieve this prestigious degree at RIT. Finally, I would like to thank RIT for giving me an opportunity to be a part of this family.

## LIST OF ABBREVIATIONS

| | |
|---|---|
| SPi | Secant Plane Index |
| TSi | Tangent Sphere Index |
| SPM | Secant Plane Method |
| TSM | Tangent Sphere Method |
| GC | Global Centroid |
| LC | Local Centroid |
| LBS | Ligand Binding Site |
| 3D | Three Dimensional |
| RSP | Rod Shaped Protein |
| PDB | Protein Data Bank |
| $\alpha$ | Alpha |
| $\beta$ | Beta |
| HTML | Hyper Text Markup Language |
| PHP | Hypertext Preprocessor |
| CSS | Cascading Style Sheets |
| GUI | Graphical User Interface |
| KB | Kilo Bytes |
| | |

# LIST OF FIGURES

## LIST OF TABLES

***Part 2: Representing Rod-Shaped Protein 3D Structures in Cylindrical Coordinates***

## ABSTRACT FOR PART 2:

Based on overall 3D structure, proteins may be grouped into two broad, general categories, namely, globular proteins or 'spheroproteins', and elongated or 'fibrous proteins'. The former comprises the significant majority. This work concerns the second general category of protein structures, namely, the fibrous or rod-shaped class of proteins (sometimes also referred to as "filamentous proteins"). Unlike an spheroprotein, a rod-shaped protein (RSP) possesses a visibly conspicuous axis along its longest dimension. To take advantage of this potential symmetry element, we decided to represent RSPs using cylindrical coordinates, $(\rho, \theta, z)$, with the z-axis as the main axis and one 'tip' of the protein at the origin. A 'tip' is defined as one of two extreme points in the protein lying along the protein axis and defining its longest dimension. To do this, we first visually identify the two tips $T_1$ and $T_2$ of the protein using appropriate graphics software, then determine their Cartesian coordinates, (h, k, l) and (m, n, o), respectively. Arbitrarily selecting $T_1$ as the tip to coincide with the origin, we translate the protein by subtracting (h, k, l) from all structural coordinates. We then find the angle $\alpha$ (in degrees) between vectors $T_1 T_2$ and the positive z-axis by computing the scalar product of vectors $T_1 T_2$ and OP where P is an arbitrary point along the positive z-axis. We typically use (0, 0, p) where p is a suitable positive number. Then we compute the cross product of the two vectors to determine the axis about which we should rotate vector $T_1 T_2$ so it will coincide with the positive z-axis. We use a matrix form of Rodrigue's formula to perform the actual rotation. Finally we apply the Cartesian to cylindrical coordinate transformation equations to the system. Thus far, we have applied the above transformation to 15 rod-shaped proteins (1QCE, 2JJ7, 2KPE, 3K2A, 3LHP, 2LOE, 2L3H, 2L1P, 1KSG, 1KSJ, 1KSH, 2KOL, 2KZG, 2KPF and 3MQC. We have also created a webserver that can take the PDB coordinate file of a rod-shaped protein and output its cylindrical coordinates based on the transformation steps described above. The URL of our web server is http://tortellini.bioinformatics.rit.edu/sxc6274/thesis2.php

# Chapter 6

## *Introduction*

### *1. Background*

Proteins have different levels of structures: primary, secondary and tertiary. Primary structure represents the sequence of amino acids, which is determined by the genetic code. Secondary structure representation consists of the amino acid residues such as α helices and β sheets. Tertiary structure is the three dimensional structure of a protein. In tertiary structures α Helices and β sheets are folded by hydrophobic interactions and are locked into place by tertiary interactions such as salt bridges, hydrogen bonds, disulfide bonds and side packing of side chains. Quaternary structure of the protein is a complex of many subunits of the protein or polypeptides that are formed by the interactions as in the tertiary structures.

In terms of gross 3D structure, the majority of the proteins are globular some are rod shaped and some are irregular in shape.

**1.1 Globular proteins:** Globular proteins are also known as spheroproteins and they are spherical in shape and are soluble in aqueous solutions. Enzymes involved in metabolic functions are mostly spheroproteins.

**1.2 Rod shaped proteins:** Rod shaped proteins are elongated in shape and, insoluble in aqueous solutions and are major component of hair, horns, nails, wool, silk etc., which play important structural role. Examples of rod shaped proteins are keratin, collagen. Most rod shaped proteins are elongated as they have repeated structures of amino acids.

## 1.3 Cartesian Coordinate System

Here we are interested in the three-dimensional structures of proteins, which is the set of coordinates that describe the position of the atoms in the protein. Cartesian coordinates in 3D are based on three mutually perpendicular axes x, y and z. Protein Cartesian Coordinate files are called PDB files and have a specific format (Berman et al. 2000).

**1.4 PDB:** Protein Data Bank (PDB) is the central repository in the world where the three dimensional structures of proteins and nucleic acids are deposited. The 3D structures are determined by using various experimental methods like X-ray diffraction, NMR, electron microscopy and other methods. The structures are deposited in a specific file format called PDB file format. The PDB file describes the position of the protein atoms as (x, y, z) coordinates in protein in three-dimensional space. It also includes the ligands that are bound to a protein as well as water molecules in the structure, if any. Our project focuses on the x, y, z coordinates of the atoms, as these represent the position of the atoms in 3D structure (Berman et al. 2000).

**1.5 Visualization of the protein 3D structure:** We used Jmol to visualize protein 3D structures. The graphic image obtained from Jmol can be clicked and rotated in every direction to have a thorough view of the protein shape (Jmol). The gross shape of proteins in JMol coluld be (1) spherical or globular (2) rod shaped or cylindrical or (3) irregularly as described previously.

# Chapter 7

## *1. Statement of the Problem*

A typical globular protein is 1CMYwhich is a composite quaternary state of human hemoglobin whose structure is shown in Figure 10. A typical rod shaped protein is 1QCE as shown in the Figure 11.

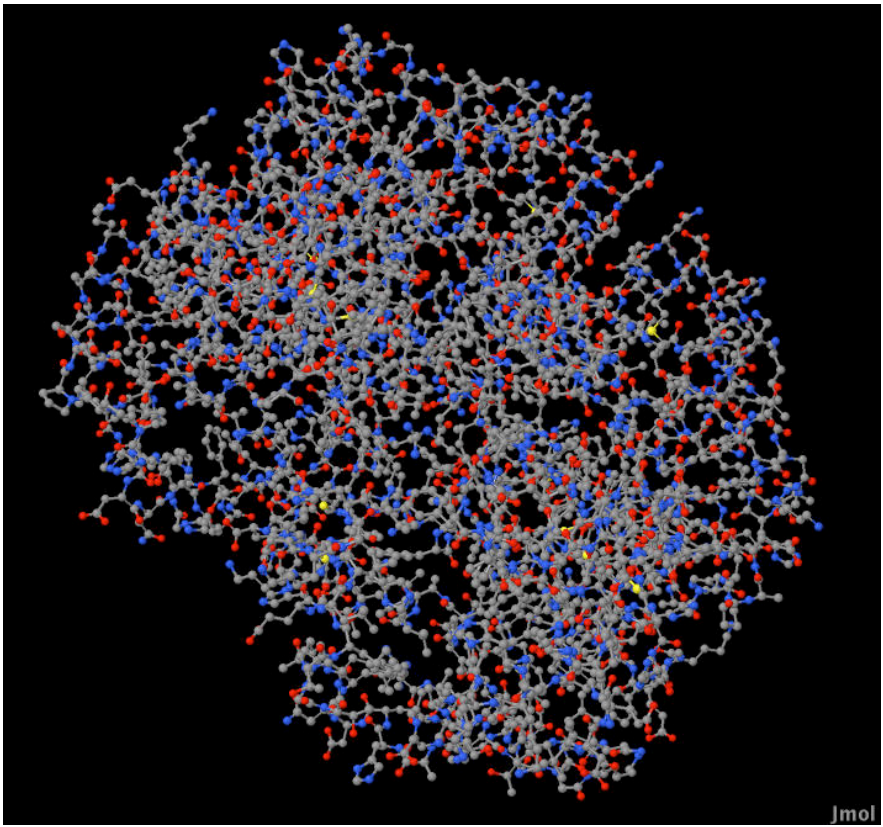

**Figure 10: structure of 1CMY (Quarternary state of human heamoglobin)**

It is evident that rod shaped or elongated proteins have a main axis running through its maximal dimension. The aim of this work is to take advantage of this inherent symmetry by using the cylindrical coordinates.

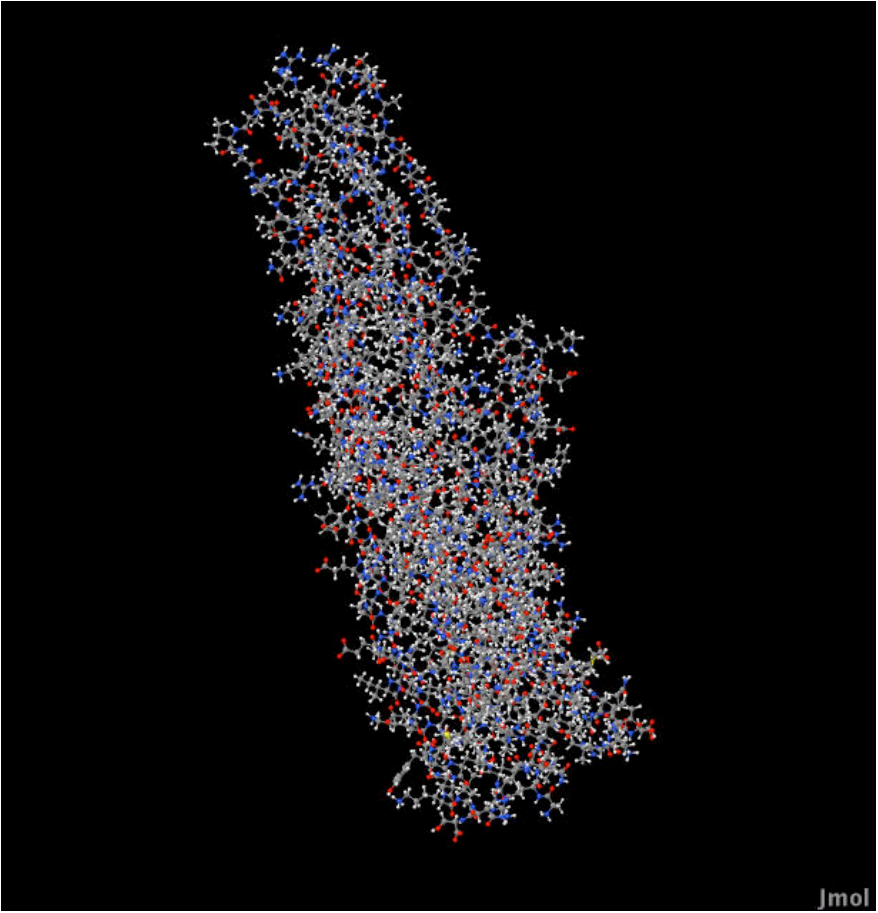

**Figure 11: Structure of 1QCE (Ectodomain of SIV GP41)**

# Chapter 8

## *1 Methods*

### 1.1 Identification of the extreme points of the rod shaped proteins

We have written a set of programs in Fortran 77 and 90 that transform the PDB (i.e., Cartesian) coordinates of a protein into Cylindrical coordinates, where one end or "tip" of the protein is at the origin and its main axis coincides with the primitive Z-axis.

To convert Cartesian coordinates to cylindrical coordinates three sequential steps have to be performed. (1) Translation (2) Rotation (3) Transformation. The extreme points, or "tips" of the rod-shaped protein are identified visually using graphical software, JMol. This step is not yet automated. Cylindrical proteins are translated, rotated and transformed based on the tip. The extreme points are the two points at the two ends, of the rod-shaped protein along its maximal dimension. Here we designate them as $T_1$ and $T_2$ points.

### 1.2 Translation: Translation preserves the distances and directions between the points in a plane. In translation fixed number of points is added to the Cartesian coordinates for every point in the plane. Translation moves all the points in the PDB in the same direction with the same distance. If (h, j, k) are the Cartesian coordinates of one of the tip points of the rod-shaped proteins, $T_1$, then translation on every point in the protein is achieved by the equation:

$(x`, y`, z`) = (x-h, y-j, z-k)$

**1.3 Rotation:** Rotation in 3D is the motion of a rigid body around a fixed rotation axis. After translation, we need to rotate the rod shaped protein along a rotation axis (a line in 3D) so that the protein's main axis is coincident with the positive Z-axis. We used two pieces of information to effect this rotation. First, the rotation axis needs to be identified. Rotation axis is determined by doing the cross product of vectors $T_1 T_2$ and OP where P is a conventional point along the positive Z-axis. Second, the amount of rotation (in degrees) must be determined. The amount of rotation is determined by doing the dot product of the same vectors $T_1 T_2$ and OP. 'Rodrigues' formula is used to perform the actual rotation (see appendix).

**1.4 Transformation:** The transformation step is the conversion of the rotated Cartesian coordinates to cylindrical coordinates. Once the rod-shaped protein has been rotated and is along the positive Z-axis, it can be transformed from Cartesian to Cylindrical coordinates. In this transformation, standard equations are used (see appendix).

In summary, the programs for the three steps are written in FORTRAN.

a) Initially a PDB file is given as the input for the translation program. The program outputs a translated PDB file.

b) The next step is to rotate the translated file. Using Rotation program the translated PDB file is converted to rotated PDB file.

c) The final step is the transformation of the coordinates by inputting rotated PDB file and it outputs transformed PDB file, which has cylindrical coordinate, RHO, PHI, Z (see results).

A shell script is written to apply all the three programs in a single step.

| Rod Shaped proteins |
| --- |
| 1QCE |
| 3MQC |
| 3K2A |
| 2JJ7 |
| 3LHP |
| 1DXX |
| 2L3H |
| 2L1P |
| 2KPE |
| 2KZG |
| 2KOL |
| 1KSG |
| 1KSH |
| 1KSJ |
| 3T5G |
| 3T5I |

Table 2: List of rod shaped proteins used for research

# Chapter 9

## *1 Web Server Implementation and Results*

A web server is designed using HTML, CSS, Java Script, and PHP in which user inputs a PDB file (without header and footer) of a rod-shaped protein and extreme point or "tip" of the protein in f10.4 format. The tip point of the protein is obtained by visual inspection using Jmol by clicking on the selected point. The Cylindrical coordinates web server is hosted at tortellini.bioinformatics.rit.edu/sxc6274/thesis2.php.

Figure 12 shows a screen shot of the cylindrical coordinates web server.

**Figure 12: Home page of "Cylindrical coordinates" web server**

Reset button sets the server back to original.

## 1.1 Results

**Web Server Results:** When uploaded protein file is in PDB format and the tip file in f10.4 format the user is redirected to the results page, where three links to translated file, rotated file and transformed file are displayed as in Figure 13. When the links are clicked the respective files for the protein are displayed.

**Results for the Conversion of Cartesian Coordinates to Cylindrical Coordinates**

Translated file of the uploaded protein
Rotated file of the uploaded protein
Cylindrical coordinates file of the uploaded protein

**Figure 13: Results page of "Cylindrical coordinates" webserver**

A portion of the final output file in cylindrical coordinates is shown in the Figure 14:

```
ATOM      1  N   VAL H   2      55.37927628   -70.10767365   13.83899975    CYL (RHO, PHI, Z) in degrees
ATOM      2  CA  VAL H   2      55.90459442   -70.72850800   12.58100035    CYL (RHO, PHI, Z) in degrees
ATOM      3  C   VAL H   2      57.31018448   -71.31623077   12.80200005    CYL (RHO, PHI, Z) in degrees
ATOM      4  O   VAL H   2      58.22136307   -70.61863708   13.25300026    CYL (RHO, PHI, Z) in degrees
ATOM      5  CB  VAL H   2      55.95610046   -69.65103912   11.43900013    CYL (RHO, PHI, Z) in degrees
ATOM      6  CG1 VAL H   2      56.45404053   -70.27346039   10.12399960    CYL (RHO, PHI, Z) in degrees
ATOM      7  CG2 VAL H   2      54.58815765   -68.97824097   11.23499966    CYL (RHO, PHI, Z) in degrees
ATOM      8  N   GLN H   3      57.50953293   -72.59029388   12.47799969    CYL (RHO, PHI, Z) in degrees
ATOM      9  CA  GLN H   3      58.82903671   -73.18927002   12.61800003    CYL (RHO, PHI, Z) in degrees
ATOM     10  C   GLN H   3      59.27198029   -73.82250214   11.31299973    CYL (RHO, PHI, Z) in degrees
ATOM     11  O   GLN H   3      58.53273392   -74.57738495   10.69699955    CYL (RHO, PHI, Z) in degrees
ATOM     12  CB  GLN H   3      58.82223511   -74.19708252   13.74400043    CYL (RHO, PHI, Z) in degrees
ATOM     13  CG  GLN H   3      60.19552612   -74.51337433   14.30099964    CYL (RHO, PHI, Z) in degrees
ATOM     14  CD  GLN H   3      60.13618851   -75.31579590   15.57400036    CYL (RHO, PHI, Z) in degrees
ATOM     15  OE1 GLN H   3      59.04722214   -75.63849640   16.07900047    CYL (RHO, PHI, Z) in degrees
ATOM     16  NE2 GLN H   3      61.31660461   -75.64805603   16.10000038    CYL (RHO, PHI, Z) in degrees
ATOM     17  N   LEU H   4      60.49272537   -73.54133606   10.89200020    CYL (RHO, PHI, Z) in degrees
ATOM     18  CA  LEU H   4      61.05133820   -74.13568878    9.71300030    CYL (RHO, PHI, Z) in degrees
ATOM     19  C   LEU H   4      62.16654587   -74.97412109   10.18400002    CYL (RHO, PHI, Z) in degrees
ATOM     20  O   LEU H   4      62.98040390   -74.60643768   11.00199986    CYL (RHO, PHI, Z) in degrees
ATOM     21  CB  LEU H   4      61.57289505   -73.16161346    8.71599960    CYL (RHO, PHI, Z) in degrees
ATOM     22  CG  LEU H   4      60.51953888   -72.19555664    8.25199986    CYL (RHO, PHI, Z) in degrees
ATOM     23  CD1 LEU H   4      61.18471527   -71.04237366    7.64599991    CYL (RHO, PHI, Z) in degrees
ATOM     24  CD2 LEU H   4      59.54490280   -72.79688263    7.30200005    CYL (RHO, PHI, Z) in degrees
ATOM     25  N   VAL H   5      62.21055603   -76.09688568    9.66800022    CYL (RHO, PHI, Z) in degrees
ATOM     26  CA  VAL H   5      63.28472900   -76.93677521    9.99800014    CYL (RHO, PHI, Z) in degrees
ATOM     27  C   VAL H   5      63.96201706   -77.38520813    8.72500038    CYL (RHO, PHI, Z) in degrees
ATOM     28  O   VAL H   5      63.29619598   -77.83985138    7.81899977    CYL (RHO, PHI, Z) in degrees
ATOM     29  CB  VAL H   5      62.78926086   -78.00804901   10.84300041    CYL (RHO, PHI, Z) in degrees
ATOM     30  CG1 VAL H   5      62.52966690   -77.57872009   12.27600002    CYL (RHO, PHI, Z) in degrees
ATOM     31  CG2 VAL H   5      61.54953003   -78.61776733   10.22599983    CYL (RHO, PHI, Z) in degrees
ATOM     32  N   GLN H   6      65.28579712   -77.25396729    9.66300011    CYL (RHO, PHI, Z) in degrees
ATOM     33  CA  GLN H   6      66.04741669   -77.50469971    7.45100021    CYL (RHO, PHI, Z) in degrees
ATOM     34  C   GLN H   6      66.77519989   -78.65742493    7.51599979    CYL (RHO, PHI, Z) in degrees
```

**Figure 14: Sample Result file showing cylindrical coordinates**

## 1.2 Validation of the Web Server: Server side and client side validations are

incorporated into the web server to prompt the user.

**Client side validation:** It includes generation of an alert button, when the user hits the submit button before uploading the files. An alert button is displayed as shown in the Figure 15. This prompts the user to upload a protein file and tip before clicking the submit button.

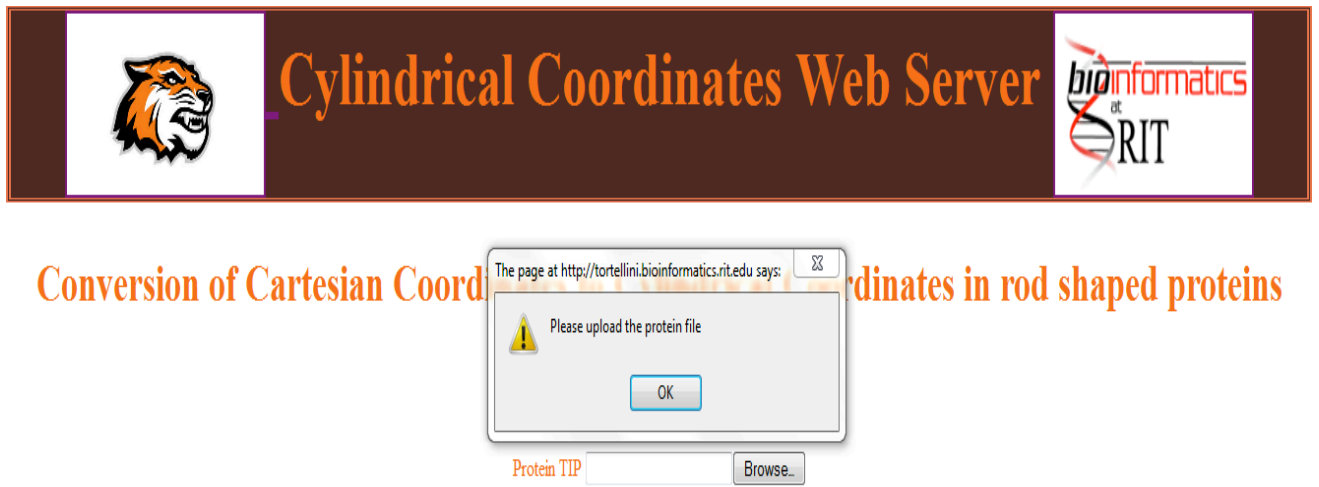

**Server side validation:** It includes checking the format of the uploaded files, it prompts the user if it is not in a right PDB format. It also checks for the size of the uploaded file. The size of the uploaded PDB file must be less than 10,000 KB and the other input file, protein tip, must be in f10.4 format.

### 1.3 Figures explaining steps of conversion

The steps of conversion from Cartesian to Cylindrical coordinates for two prominent rod shaped proteins are depicted as in the figures below:

*Figures 16,17,18 and 19 describe transformation steps for the protein 3MQC (Crystal Structure of Ectodomain of BST-2/Tetherin/CD317).*

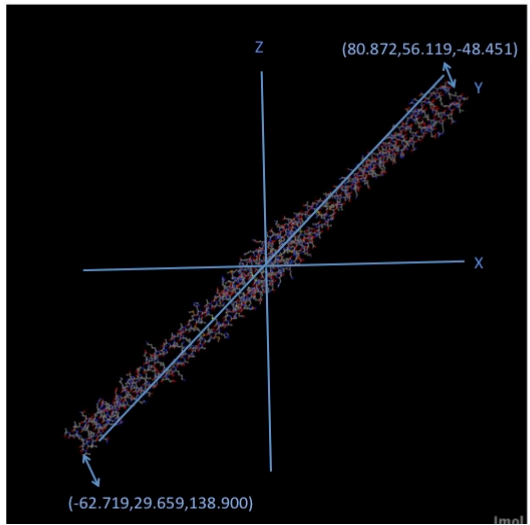

**Figure 16:3MQC with tips identified**

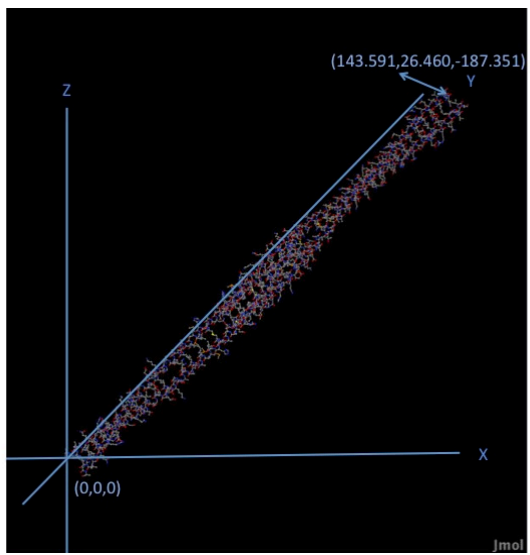

**Figure 17: 3MQC after Translation**

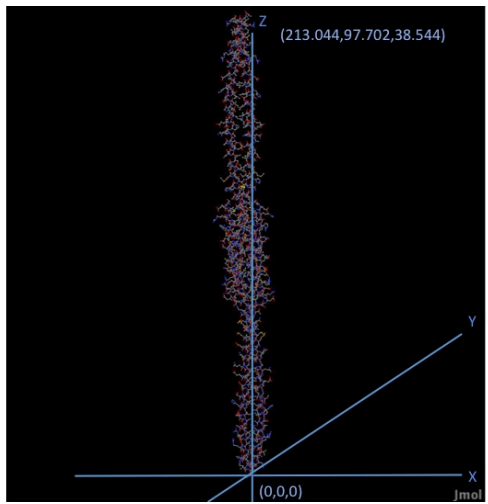

**Figure 18: 3MQC after rotation**

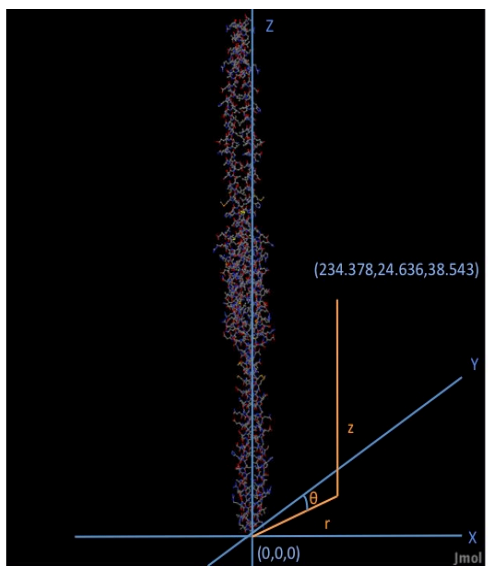

**Figure 19: 3MQC after transformation**

*Figures 20,21,22 and 23 describe transformation steps for the protein 2JJ7 (crystal structure of HLYIIR mutant protein).*

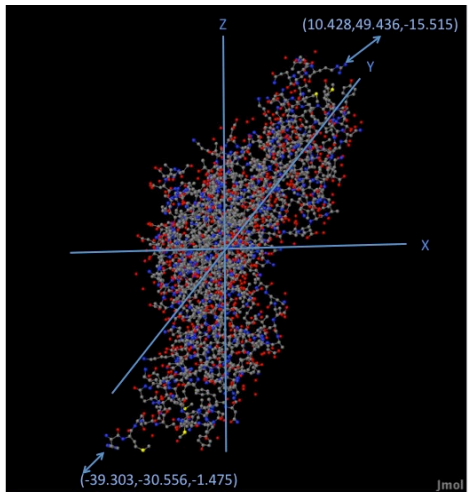

**Figure 20: 2JJ7 with tips identified**

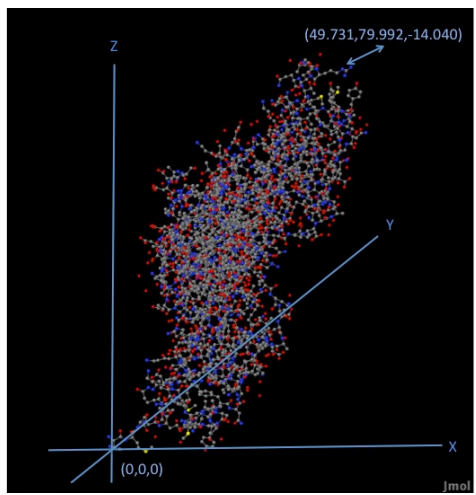

**Figure 21: 2JJ7 after translation**

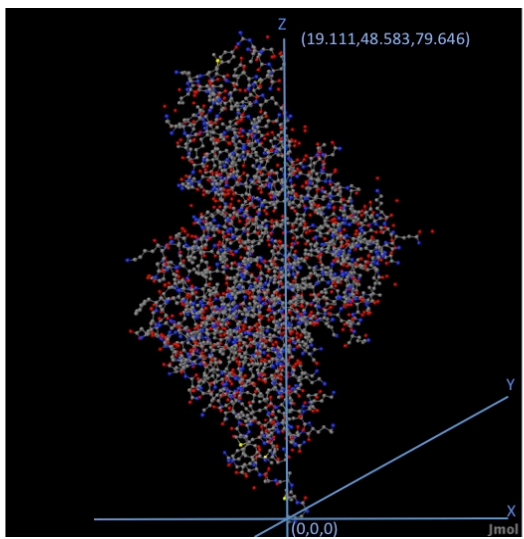

**Figure 22: 2JJ7 after rotation**

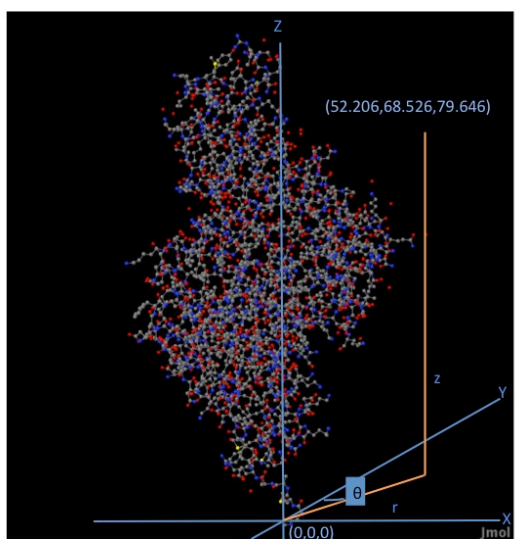

**Figure 23: 2JJ7 after transformation**

## 1.4 Conclusion:

To the best of our knowledge, this work is the first time that rod shaped proteins, or any protein for that matter have been represented in cylindrical coordinates. Current visualization tools like JMol, Pymol does not visualize the cylindrical coordinates, as this is first method that converts Cartesian to cylindrical coordinates.

# References:

1. Ask the CSS Guy. (2011). Form field hints with CSS and JavaScript. Retrieved March 20 2011 from http://www.askthecssguy.com/2007/03/form_field_hints_with_css_and.html

2. Berman, H. M., Westbrook, J., Feng, Z., Gilliland, G., Bhat, T. N., Weissig, H., Bourne, P. E. (2000). The protein data bank. *Nucleic Acids Research, 28*(1), 235-242.

3. Binding Site. (2011). Binding site of dihydrofolate reductase (3dfr). Retrieved July 11 2011, from http://www.biochem.ucl.ac.uk/~roman/surfnet/examples/3dfr_site.html

4. Brady, G. P.,Jr, & Stouten, P. F. (2000). Fast prediction and visualization of protein binding pockets with PASS. *Journal of Computer-Aided Molecular Design, 14*(4), 383-401.

5. Coleman, R. G., & Sharp, K. A. (2006). Travel depth, a new shape descriptor for macromolecules: Application to ligand binding. *Journal of Molecular Biology, 362*(3), 441-458. doi:10.1016/j.jmb.2006.07.022

6. Hendlich, M., Rippmann, F., & Barnickel, G. (1997). LIGSITE: Automatic and efficient detection of potential small molecule-binding sites in proteins. *Journal of Molecular Graphics & Modelling, 15*(6), 359-63, 389.

7. Introductory Biology, Cornell University. (2011) Fibrous vs Globular Proteins.Retrieved July 15 2011 from http://www.biog11051106.org/demos/105/unit1/fibrous_v_glob.html

8. Jmol. (2011). Jmol: an open-source Java viewer for chemical structures in 3D. Retrieved July 15, 2011 from http://www.jmol.org

9. Laskowski, R. A., Luscombe, N. M., Swindells, M. B., & Thornton, J. M. (1996). Protein clefts in molecular recognition and function. *Protein Science : A Publication of the Protein Society, 5*(12), 2438-2452. doi:10.1002/pro.5560051206

10. Laurie, A. T., & Jackson, R. M. (2005). Q-SiteFinder: An energy-based method for the prediction of protein-ligand binding sites. *Bioinformatics (Oxford, England), 21*(9), 1908-1916. doi:10.1093/bioinformatics/bti315

11. PHP. (2011). PHP Tutorial. Retrieved July 15 2011 from http://www.w3schools.com/php/

12. Pocket-Finder Pocket Detection. (2011). Detection of Pockets using Pocket Finder Retrieved July 15 2011 from http://www.modelling.leeds.ac.uk/pocketfinder/

13. Shen, S., & Tuszynski, A. (2008). *Theory and Mathematical methods in Bioinformatics*. Berling: Germany.

14. Sheth, V.N., (2009). Visualization of Protein 3D Structures in Reduced Representation with Simultaneous Display of Intra and Inter-molecular Interactions (Master's Thesis). Retrieved March 15 2011from http://rc.rit.edu/docs/sheth-thesis-10.09.pdf

15. thesitewizard.com. (2011). Form Input Validation JavaScript. Retrieved March 20 2011 from http://www.thesitewizard.com/archive/validation.shtml

16. Voet D, Voet J.G. *Biochemistry*. New York: John Wiley and Sons, 1990. Print.

17. Weisel, M., Proschak, E., & Schneider, G. (2007). PocketPicker: Analysis of ligand binding-sites with shape descriptors. *Chemistry Central Journal, 1*, 7. doi:10.1186/1752-153X-1-7

```
{
echo "Error: Only text files and below the size of 100,000 KB are accepted";
}
?>
```

## PART 2

### Wrapper Script

```
cp tip filea; cp protein fileb; ./translate.x;
cp fileo filei; ./Zrotation.x;
cp fileo filei; ./cart2cyl_v2_deg.x;
```

### Front end Code

```html
<html>
<head>
<style type="text/css">
h1
{
    background-color: #513127;
    border-style: double;
    border-width: 3px;
    border-left-width:5px;
    border-right-width:5px;
    border-color: #E67451;
    margin-top: -8.5px;
    margin-right: -5px;
    margin-left: -5px;
    <!--margin: 0.5em;-->
    padding:4em;
    font-size:50px;
    }

h2
{
font-size:34px;
}
p
{
```

```
font-family:Verdana;
font-size:15px;
}

</style>

<!-- Reference:Form Input Validation
(http://www.thesitewizard.com/archive/validation.shtml) -->
    <script type="text/javascript" language="javascript">
    function validate_form (form )
    {
       if( form.protein.value == "")
       {
          alert("Please upload the protein file");
          form.protein.focus();
          return false;
       }
       return true;
    }
    function trim(str)
    {
       return str.replace(/^\s+|\s+$/g,'');
    }
}

</script>
<!-- Reference:Form Input Validation  ends here -->
</head>

<body>
<center>
<font color="#F36E21">
 <h1> <a href="http://rit.edu">
<img align=middle src="tiger.jpg" alt="Rochester Institute of Technology" height="120"
width="200"> </img> </a> Cylindrical Coordinates Web Server <a
href="http://www.bioinformatics.rit.edu">
<img align=middle src="bioinfologo.gif" alt="Bioinformatics at RIT" height="120"
width="200"> </img> </a> </h1>
<!-- </font>
<font color="#F36E21"> -->
<h2>
Conversion of Cartesian Coordinates to Cylindrical Coordinates in rod shaped proteins
</h2>
<br />
<form action="result2.php" name="file_form" method="post" enctype="multipart/form-
data" onSubmit="return validate_form (this );"  >
```

Protein File <input type="file" name="protein" input type="hidden" name="MAX_FILE_SIZE" value="1000000" /> <br /> <br />
Protein TIP <input type="file" name="tipper" input type="hidden" name="MAX_FILE_SIZE" value="300" />

<br />
 <br /> <br />
<input type="submit" value="Submit"  />
<input type="reset" value="Reset" />

</form>
</center>
<br />
<br />
<br />
<p align="left"> <b> <i> For your queries contact us: <br />
Srujana Cheguri - sxc6274@rit.edu<br />
Dr.Vicente Reyes - vmrsbi@rit.edu  </b> </i>
<br /> <br />
<p><b>Please check our other Web Server <a
href="http://tortellini.bioinformatics.rit.edu/sxc6274/thesis1.php"></b><br />Ligand
Burial Site Depth Determination</a></p>
</p>

</font>
</body>
</html>

**Back End Code**

<html>
   <head>
   <title>
   Results Page
   </title>
<style type="text/css">
h1
   {
   background-color: #513127;
   border-style: solid;
   border-width: 3px;
   border-left-width:5px;
   border-right-width:5px;
   border-color: #E67451;
   margin-top: -8.5px;

```
    margin-right: -5px;
    margin-left: -5px;
    <!--margin: 0.5em;-->
    padding:4em;
    font-size:50px;
    }
h2
{
font-size:27px;
}
p
{
font-family:Verdana;
font-size:15px;
}
</style>
</head>
<body>
<center>
<font color="#F36E21">
 <h1> <a href="http://rit.edu">
<img align=middle src="tiger.jpg" alt="Rochester Institute of Technology" height="120"
width="200"> </img> </a> Cylindrical Coordinates Web Server <a
href="http://www.bioinformatics.rit.edu">
<img align=middle src="bioinfologo.gif" alt="Bioinformatics at RIT" height="120"
width="200"> </img> </a> </h1>
<!-- </font>
<font color="#F36E21" -->
<h2> Results for the Conversion of Cartesian Coordinates to Cylindrical
Coordinates</h2>
<!-- <p>Click here for the results <a
href="http://tortellini.bioinformatics.rit.edu/sxc6274/files/$dirname/fileo"> result </a>
</p> -->
</font>
<center>

</body>
</html>
<?php
#Reporting the simple errors
// Report simple running errors
ini_set('error_reporting', E_ALL ^ E_NOTICE);

// Set the display_errors directive to OFF
ini_set('display_errors', 0);
```

```php
// Log errors to the web server's error log
ini_set('log_errors', 1);

// Destinations
define("ADMIN_EMAIL", "sxc6274@rit.edu");
define("LOG_FILE", "/home/sxc6274/public_html/error2.log");

// Destination types
 /*define("DEST_EMAIL", "1");
  */
define("DEST_LOGFILE", "3");
/* Examples */

// Send an e-mail to the administrator
error_log("Fix me!", DEST_EMAIL, ADMIN_EMAIL);

// Write the error to our log file
error_log("Error", DEST_LOGFILE, LOG_FILE);

function my_error_handler($errno, $errstr, $errfile, $errline)
    {
    switch ($errno) {
       case E_USER_ERROR:
           // Send an e-mail to the administrator
           error_log("Error: $errstr \n Fatal error on line $errline in file $errfile \n",
DEST_EMAIL, ADMIN_EMAIL);

           // Write the error to our log file
           error_log("Error: $errstr \n Fatal error on line $errline in file $errfile \n",
DEST_LOGFILE, LOG_FILE);
           break;

       case E_USER_WARNING:
           // Write the error to our log file
           error_log("Warning: $errstr \n in $errfile on line $errline \n", DEST_LOGFILE,
LOG_FILE);
           break;

       case E_USER_NOTICE:
           // Write the error to our log file
           error_log("Notice: $errstr \n in $errfile on line $errline \n", DEST_LOGFILE,
LOG_FILE);
           break;
```

```php
        default:
            // Write the error to our log file
            error_log("Unknown error [#$errno]: $errstr \n in $errfile on line $errline \n",
DEST_LOGFILE, LOG_FILE);
        break;
    }
    // Don't execute PHP's internal error handler
    return TRUE;
    }
// Use set_error_handler() to tell PHP to use our method
$old_error_handler = set_error_handler("my_error_handler");
#include('/home/sxc6274/config.php');
#http://myphpform.com/validating-forms.php
$file1 = input_val($_POST["protein"]);
chmod("$file1",0777);
$file2 = input_val($_POST["tipper"]);
chmod("$file2",0777);
function input_val($data)
    {
    $data = trim($data);
    $data = stripslashes($data);
    $data = htmlspecialchars($data);
    return $data;
    }
$filename1=basename($_FILES["protein"]["name"]);
#chmod("$filename1",0777);
$filename2=basename($_FILES["tipper"]["name"]);
#chmod("$filename2",0777);
#echo $filename1;
$findkey1='ATOM';
$findkey2='HETATM';
$ext1=substr($filename1,strpos($filename1,'.')+1);
$ext2=substr($filename2,strpos($filename2,'.')+1);

$target_path = "/home/sxc6274/public_html/files/";
$newname1 = $target_path.$filename1;
#echo "$newname1 <br>\n";
#chmod("$newname1",0777);
$newname2 = $target_path.$filename2;
#echo "$newname2 <br>\n";
#chmod("$newname2",0777);
$dirname=substr($filename1,0,4);
#echo "$dirname <br> \n";
if((($ext1=="txt") && ($_FILES["protein"]["size"] < 1000000)) && (($ext2=="txt") &&
($_FILES["tipper"]["size"] < 500)))
```

```php
{
    if((is_uploaded_file($_FILES['protein']['tmp_name']))&&(is_uploaded_file($_FILES['tipper']['tmp_name'])))
    {
    $handle1 = fopen($_FILES['protein']['tmp_name'], "r");
        $handle2 = fopen($_FILES['tipper']['tmp_name'], "r");
        $line1=fgets($handle1);
        $line2=fgets($handle2);
        #echo "$line1 <br> \n";
        #echo "$line2 <br> \n";
        $pos1=strpos($line1,'ATOM');
        #echo "$pos1 for pos1 <br> \n";
        #$pos2=strpos($line1,'HETATOM');
        #echo "$pos2 for pos2 <br> \n";
    if($pos1===FALSE)
        {
            echo "Not a PDB file <br> \n";
            exit(1);
        }

        else
        {
            while(!feof($handle1) && (!feof($handle2)))
            {
                $line1=fgets($handle1);
                $line2=fgets($handle2);

                if($pos1 == 0)
                {
                    $nc=preg_split("/\s+/",$line1);
                    $x=sizeof($nc);
                    #echo "$x <br> \n";

                    if (sizeof($nc)== 13 || 12)
                    {
                        break;
                    }
                    else
                    {
                        echo "Not in PDB Format <br> \n";
                        exit(1);
                    }
                }
            }
        }
```

```
}

else
{
    echo "Not uploaded right type of files <br> \n";
    exit(1);
}if((!file_exists($newname1))&&(!file_exists($newname2)))
{
    @system("rm -r /home/sxc6274/public_html/files/$dirname");
    #echo "entered file_exists if loop <br> \n";

    $rval1= move_uploaded_file($_FILES['protein']['tmp_name'],$newname1);
    $rval2 = move_uploaded_file($_FILES['tipper']['tmp_name'],$newname2);
    $rval = $rval1 && $rval2;
    #echo "$rval1 <br>\n";
    #echo "$rval2 <br>\n";
    #echo "$rval <br>\n";

    if($rval)
    {
        #echo "succesfully entered the loop of move uploaded file";
        @system ("mkdir /home/sxc6274/public_html/files/$dirname");
        @system("mv $newname1 /home/sxc6274/public_html/files/$dirname/protein");
        @system("mv $newname2 /home/sxc6274/public_html/files/$dirname/tip");
    @system("cp /home/sxc6274/public_html/files/scriptthesis2.sh
/home/sxc6274/public_html/files/$dirname");
        @system("cp /home/sxc6274/public_html/files/pre_process.f
/home/sxc6274/public_html/files/$dirname");
        @system("cp /home/sxc6274/public_html/files/pre_process.x
/home/sxc6274/public_html/files/$dirname");
        @system("cp /home/sxc6274/public_html/files/Zrotation.f
/home/sxc6274/public_html/files/$dirname");
        @system("cp /home/sxc6274/public_html/files/Zrotation.x
/home/sxc6274/public_html/files/$dirname");
        @system("cp /home/sxc6274/public_html/files/translate.f
/home/sxc6274/public_html/files/$dirname");
        @system("cp /home/sxc6274/public_html/files/translate.x
/home/sxc6274/public_html/files/$dirname");
        @system("cp /home/sxc6274/public_html/files/cart2cyl_v2_deg.f90
/home/sxc6274/public_html/files/$dirname");
        @system("cp /home/sxc6274/public_html/files/cart2cyl_v2_deg.x
/home/sxc6274/public_html/files/$dirname");
            chdir ("/home/sxc6274/public_html/files/$dirname");
        #chmod ("/home/sxc6274/public_html/files/$dirname",0777);
        #system('pwd');
```

```php
            @system("./scriptthesis2.sh");
    #$out=file_get_contents('fileo');
            #echo "$out";
            #chmod("$out",0777);
            #system("cat fileo");
            /*$lines = 0;
            if ($fh = fopen('fileo', 'r')) {
            while (!feof($fh)) {
            if (fgets($fh)) {

             $linex=fgets($fh);
             echo "$linex <br> \n";
             $lines++;
              }
              }
              }
            echo $lines; // line count*/
            #echo "$dirname  dirname<br> \n";

            if(file_exists('fileo'))
              {
                    chdir ("/home/sxc6274/public_html/files/$dirname/");
                    #system('pwd');
                    echo "You can view your result by clicking the link below.<br>";
                    $link="http://tortellini.bioinformatics.rit.edu/sxc6274/files/$dirname/fileo"
;
                    #echo "copy and paste this  $link <br> \n";
                    #print ('<a href="$link"> results </a>');
                    echo " <a href =
'http://tortellini.bioinformatics.rit.edu/sxc6274/files/$dirname/fileo'> Results </a> ";

              }
            #@system("rm /home/sxc6274/public_html/files/$dirname/fileo");

        }

      }
    else
      {
        echo "Error: File ".$_FILES["protein"]["tmp_name"] ." and
".$_FILES["tipper"]["tmp_name"] ." already exists";
      }
```

```
}

else
{

    echo "Error: Only text files and below the size of 100,000 KB are accepted";
}

?>
```